\documentclass[runningheads]{llncsMod}
\usepackage{amsmath,amssymb}
\usepackage{xspace}
\usepackage[usenames,dvipsnames,table]{xcolor}
\usepackage{wrapfig,graphicx}
\DeclareGraphicsExtensions{.pdf,.png}
\graphicspath{{figs/}{plots/}}
\usepackage{subcaption}
\usepackage{alltt}
\usepackage{tikz}
\usepackage{bm}
\usepackage[inline]{enumitem}

\usepackage{booktabs}
\usepackage{multirow}
\usepackage{siunitx}
\usepackage{placeins}

\usepackage{mathtools}
\DeclarePairedDelimiter\set{\lbrace}{\rbrace}
\DeclarePairedDelimiter\abs{\lvert}{\rvert}
\DeclarePairedDelimiter\ceil{\lceil}{\rceil}
\def\Oh{\ensuremath{\mathcal{O}}}
\def\cC{\ensuremath{\mathcal{C}}}
\def\cS{\ensuremath{\mathcal{S}}}
\def\cT{\ensuremath{\mathcal{T}}}
\DeclareMathOperator{\prob}{Pr}

\usepackage{url}
\usepackage[round]{natbib}
\definecolor{linkblue}{rgb}{0.198,0.198,0.5392}
\usepackage[colorlinks=true,
    linkcolor=linkblue,
    anchorcolor=linkblue,
    citecolor=linkblue,
    filecolor=linkblue,
    menucolor=linkblue,
    urlcolor=linkblue,
    bookmarks=true,
    bookmarksopen=true,
    bookmarksopenlevel=2,
    bookmarksnumbered=true,
    hyperindex=true,
    plainpages=false,
    pdfpagelabels=true
]{hyperref}
\usepackage[capitalise,nameinlink]{cleveref}

\renewcommand{\orcidID}[1]{\href{https://orcid.org/#1}{\includegraphics[scale=.03]{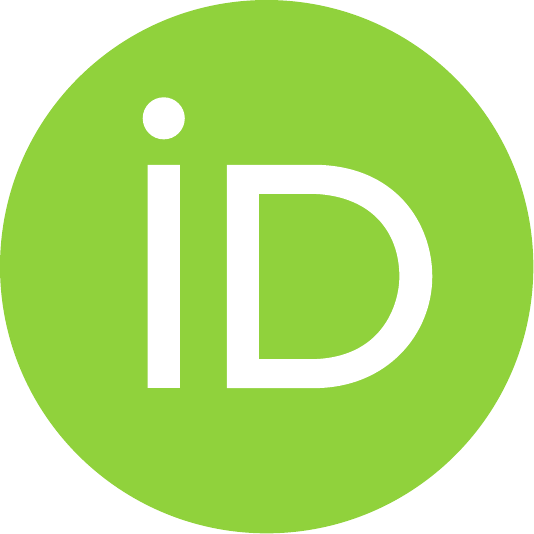}}}

\title{Bayesian Credible Sets for Phylogenetic Tree Topologies with Applications to Coverage Analysis and Cross-Model Comparison}
\titlerunning{Bayesian Credible Sets for Phylogenetic Tree Topologies}
\author{Jonathan~Klawitter$^{\star}$\orcidID{0000-0001-8917-5269}
\and Alexei~J.~Drummond\orcidID{0000-0003-4454-2576}}
\authorrunning{J. Klawitter and A.J. Drummond}
\institute{University of Auckland, Aotearoa/New Zealand\\
$^\star$ \href{mailto:jo[dot]klawitter[at]gmail[dot]com}{jo.klawitter[at]gmail.com}}

\begin{document}

\maketitle

\pdfbookmark[1]{Abstract}{Abstract}
\begin{abstract}
Credible intervals and credible sets, such as highest posterior density (HPD) intervals, form an integral statistical tool in Bayesian phylogenetics, both for phylogenetic analyses and for development. Readily available for continuous parameters such as base frequencies and clock rates, the vast and complex space of tree topologies poses significant challenges for defining analogous credible sets. Traditional frequency-based approaches are inadequate for diffuse posteriors where sampled trees are often unique. To address this, we introduce novel and efficient methods for estimating the credible level of individual tree topologies using tractable tree distributions, specifically Conditional Clade Distribution (CCD). Furthermore, we propose a new concept called $\alpha$ credible CCD, which encapsulates a CCD whose trees collectively make up $\alpha$ probability. We present algorithms to compute these credible CCDs efficiently and to determine credible levels of tree topologies as well as of subtrees.
We evaluate the accuracy of these credible set methods leveraging simulated and real datasets. Furthermore, to demonstrate the utility of our methods, we use well-calibrated simulation studies to evaluate the performance of different CCD models. In particular, we show how the credible set methods can be used to conduct rank-uniformity validation and produce Empirical Cumulative Distribution Function (ECDF) plots, supplementing standard coverage analyses for continuous parameters. 
\end{abstract}

\section{Introduction}  
Bayesian phylogenetic inference seeks to estimate the posterior distribution of phylogenetic trees and model parameters from sequence data under a specified evolutionary model. The primary computational method is Markov chain Monte Carlo (MCMC), which iteratively samples the state space~\citep{ronquist12mrbayes,hoehna16revbayes,suchard2018bayesian,bouckaert2019beast}. For continuous parameters (e.g.\ kappa, base frequencies, molecular clock rate, population size), the posterior distributions can be readily summarised by considering statistics of the marginal distribution of the parameter of interest from the samples obtained by MCMC. These include estimates of mean and variance as well as \emph{credible intervals}, e.g.\ the 95\% highest posterior density (HPD) interval. Credible intervals are a fundamental statistical tool to quantify uncertainty, allowing researchers to assess the reliability of parameter estimates, compare models, and test hypotheses. 

One of the main challenges in phylogenetic inference is that the tree topology (often the main parameter of interest) is a discrete parameter with a vast and complex state space that grows super-exponentially with the number of taxa~\citep{billera01,gavryushkin2016space}. For datasets with diffuse posteriors -- indicated, for example, by a high phylogenetic entropy~\citep{lewis16estimating,klawitter25} -- the MCMC sample misses the majority of topologies with non-negligible probability in the target posterior distribution, often including the mode. Moreover, most if not all sampled topologies are unique~\citep{baele24,berling25tractable}. For discrete parameters, an \emph{$\alpha$ credible set} (analogous to a credible interval) may be defined as a set of elements that accumulate $\alpha$ of the probability mass, e.g.\ for $\alpha = 0.95$. Given the challenges of treespace, a frequency-based credible set on the sample is impractical and incomplete for all but trivially small datasets~\citep{magee2023trustworthy}. While credible intervals are widely used for continuous parameters, a lack of suitable methods hinders their use for phylogenetic trees~\citep{whidden20}.

The idea of Bayesian credible sets for tree topologies dates back to the earliest MCMC methods in phylogenetics. \citet{mau1997phylogenetic,mau1999bayesian} first explicitly constructed credible regions by ranking sampled topologies by their MCMC visit frequency, and \citet{yang1997bayesian} similarly used MCMC frequencies to estimate posterior probabilities of individual trees. \citet{huelsenbeck2004frequentist} evaluated the frequentist coverage of 95\% credible sets constructed this way under simulation, while \citet{drummond2006relaxed} used frequency-based credible sets to assess the accuracy and precision of relaxed clock dating methods. In all cases, the credible set is constructed by the same procedure: rank sampled trees by their MCMC frequency and accumulate them until the desired probability mass is reached. More recently, \citet{jenningsshaffer24} sought to identify high posterior density regions of tree space using systematic NNI-based expansion of sDAGs (explained below), but their evaluation still relies on frequency-based credible sets from very long MCMC runs as ground truth.

A parallel tradition in the frequentist framework constructs \emph{confidence sets} of tree topologies by inverting hypothesis tests. The Kishino--Hasegawa (KH) test \citep{kishino1989evaluation} compares the log-likelihood difference between two \emph{a priori} specified topologies, while the Shimodaira--Hasegawa (SH) test \citep{shimodaira1999multiple} corrects for the selection bias that arises when the maximum likelihood tree is among the candidates (see \citealt{goldman2000likelihood} for a review). Shimodaira's approximately unbiased (AU) test \citep{shimodaira2002approximately} provides a further refinement using multi-scale bootstrap. A confidence set is formed as the collection of candidate topologies not rejected at level $\alpha$. \citet{willis2019confidence} takes a different geometric approach, constructing confidence sets for metric trees in Billera--Holmes--Vogtmann tree space \citep{billera01} using log-map projections, though this addresses trees with branch lengths rather than discrete topologies. All of these frequentist approaches differ fundamentally from our Bayesian credible sets: they do not produce a probability distribution over topologies, and the resulting confidence sets have a frequentist rather than Bayesian interpretation.

To fill this gap, we propose novel methods for generating credible sets of Bayesian phylogenetic tree topologies.

Recently, graph-based models of phylogenetic tree topology distributions have received considerable attention for their ability to compactly represent huge numbers of trees and estimate the posterior distribution \citep{hoehna12guided,larget13estimation,zhang18generalizing,zhang18variational,jun23,klawitter25,berling25tractable,jenningsshaffer24}. These models employ an independence assumption on the constituent parts of trees, that is, on clades and clade splits for rooted trees, and on splits for unrooted trees. (Here we focus on rooted trees as modern phylogenetic inference that includes time-tree priors requires them \citep{suchard2018bayesian,bouckaert2019beast}). The first such model framework is a \emph{conditional clade distribution (CCD)}, introduced by \citet{hoehna12guided} and refined by \citet{larget13estimation}. CCDs contain all tree topologies amalgamated from observed clades or clade splits and offer a trade-off between bias and variance (ease of accurate parameter estimation) based on the strength of the independence assumptions \citep{larget13estimation,berling25tractable}. CCDs have been used to compute the phylogenetic entropy and detect conflict among data partitions \citep{lewis16estimating}, for species tree–gene tree reconciliation \citep{szollosi13efficient}, and for guiding tree proposals for MCMC runs \citep{hoehna12guided}. Using a graph as underlying data structure for a CCD has led to efficient algorithms that allow us, among other tasks, to compute the mode of a CCD, providing superior tree topology point estimates \citep{berling25tractable}, and to unearth underlying skeleton distributions of a CCD with high support via detection and removal of rogue taxa \citep{klawitter25}. The model called \emph{subsplit directed acyclic graph (sDAG)} is based on a Bayesian network over clade splits (or splits for unrooted trees) \citep{zhang18generalizing,zhang18variational,jun23}. While conceptually similar to a particular CCD model (details below), sDAGs employ a different data structure. Furthermore, methods other than those directly using an MCMC sample to estimate their parameters have been studied. Choosing the best model remains an open problem as the performance depends on the sample size and dimensionality of the problem in non-trivial ways~\citep{whidden2015quantifying,zhang18generalizing,berling24automated}. 

In this paper, we first present a simple and efficient method to estimate the credible level of a phylogenetic tree topology for CCD and sDAG models. Furthermore, we introduce novel approaches allowing us to move beyond credible levels of single trees. Namely, we define a \emph{credible distribution} of a posterior tree distribution based on a CCD. The main idea is that an $\alpha$ credible CCD contains all trees with collectively at least $\alpha$ probability with the fewest clades or clade splits. We describe algorithms to efficiently compute credible CCDs and to determine a tree topology's credible level. Finally, we evaluate the accuracy and effectiveness of our methods on both simulated and empirical~datasets. 

A standard method to evaluate new methods and models for Bayesian phylogenetic inference is a \emph{well-calibrated simulation study (WCSS)} \citep{mendes2024validate}. Such a simulation study first samples parameters, including a \emph{true tree}, from a prior distribution, then simulates sequences on these parameters, and finally uses the inference algorithm on the same prior and model used for simulation to infer posterior distributions. This is usually repeated $\sim100$ times. For each continuous parameter, a \emph{coverage analysis} is used to test whether the inferential method works correctly, namely, whether the true parameter is contained in the HPD interval at the expected frequency. For example, for 100 simulations and $95\%$ HPD intervals, the 95\%-central interquantile interval for the number of simulations containing the true parameter is between 90 and 99 \citep{mendes2024validate}. Another closely related method is \emph{rank-uniformity validation (RUV)}, where the true parameter is ranked against the sample and, over all simulations, we expect the rank of the true parameter to be distributed uniformly. However, both methods cannot be applied directly to the tree topology and instead proxy parameters such as the total tree length, tree height, or tree distances are used \citep{mendes2024validate}. With the introduction of our credible sets, we enable such analyses for the tree topology beyond the narrow setting of datasets with particularly compact posteriors, where simple frequency counting already suffices. Indeed, we use these tests to advance our understanding of the varying performance of the different CCD models.  

An open source implementation is freely available in the CCD package for BEAST2\footnote{GitHub CCD package \href{https://github.com/CompEvol/CCD}{\texttt{github.com/CompEvol/CCD}}}, which also offers a tool to test containment of trees in credible~sets. 

\section{Materials and Methods} 
\label{sec:methods}
In this section, we first describe the CCD framework in general as well as the different CCD models, and then introduce the different types of credible sets in detail. Furthermore, we explain how the credible sets can be used in a coverage analysis of the tree topology. Throughout all trees are rooted trees and we write phylogenetic tree instead of phylogenetic tree topology, since we do not consider branch lengths or divergence times in this paper. For all running time analyses, we assume constant-time retrieval for hash map lookups of bit sets of length $n$ (where $n$ is the number of taxa). While we describe everything in the regular probability space, implementations might use transformations to the log-probability space to prevent numerical underflow and to improve numerical stability.

\subsection{Conditional Clade Distributions} 
\label{sec:ccd}
\begin{figure}[tbh]
  \centering
  \includegraphics{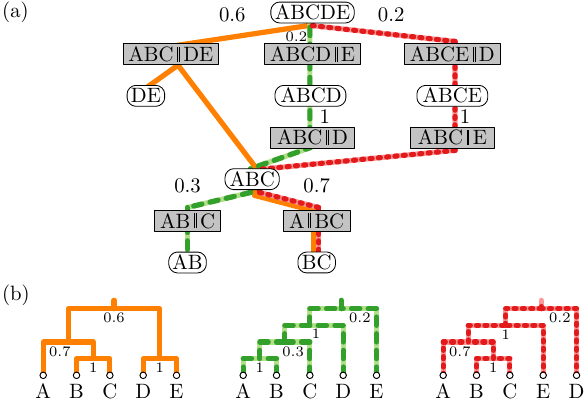}
  \caption{A CCD and three trees it contains: (a)~Truncated CCD graph (cherry splits and leaf clades omitted) is annotated with clade split probabilities.
    Only the clades ABCDE and ABC split in multiple ways. (b)~Three of the six trees represented by the CCD with probabilities $0.42$, $0.06$, and $0.14$, respectively.}
  \label{fig:ccdExample}
\end{figure}

A \emph{conditional clade distribution (CCD)} is a tree distribution that estimates the posterior over tree topologies by assuming independence between clades or clade splits. More formally, a CCD is a parametrisation of tree space, with \emph{CCD0}, \emph{CCD1}, and \emph{CCD2} expressing successively more complex models \citep{berling25tractable}. Underlying a CCD is a \emph{CCD graph}~$G$, which is a directed bipartite graph  with the two vertex sets representing the clades~$\cC(G)$ and the clade splits~$\cS(G)$; see \cref{fig:ccdExample} for an example. For a clade $C$ that is split into clades $C_1, C_2$, there is an edge from $C$ to clade split $\set{C_1, C_2}$ in~$G$ as well as an edge from $\set{C_1, C_2}$ to each of $C_1$ and $C_2$. The root of $G$ is the clade that represents the full taxon set~$X$ and the leaves of $G$ are the individual taxa in $X$. A CCD graph \emph{represents (contains)} all trees that can be constructed by starting at the root clade and then recursively picking a clade split for each clade to progress. Furthermore, each clade $C$ has a distribution on all clade splits~$\cS(C)$ of~$C$, that is, we have a probability of picking a particular clade split $S$ of~$C$. This is called the \emph{conditional clade probability (CCP)} $\prob_\pi(S)$ of $S$:
\begin{equation} \label{eq:ccp:def}
    \prob_\pi(S) = \prob(S \mid C)
\end{equation}
(This is where the name \textit{conditional clade distribution} comes from; the subscript~$\pi$ is used to indicate that this probability is conditional on the parent clade $\pi(S) = C$ of $S$.)
For a tree~$T$, let $\cS(T)$ be the set of clade splits of $T$. The probability of $T$ in the CCD is then given by the product of CCPs:
\begin{equation} \label{eq:ccp}
    \prob(T) = \prod_{S \in \cS(T)} \prob_\pi(S)
\end{equation}
For example, the left tree in \cref{fig:ccdExample} has probability $0.6 \cdot 0.7 \cdot 1 \cdot 1 = 0.42$. As \citet{larget13estimation} has shown, a CCD is a distribution on trees (that is, the probabilities of all trees in $G$ add up to one). For further details and examples, we refer to \citet{larget13estimation}, \citet{lewis16estimating}, and \citet{berling25tractable}.

Note that the number of edges of a CCD graph $G$ is linear in the number of vertices of $G$, since each clade split and each leaf has in-degree one and $G$ is~bipartite.

\paragraph{Parametrisations.}
The parameters of a CCD0 are the non-zero \emph{clade}\linebreak[4] {weights}~$w(C)$, one for each clade~$C$ in the model, and the CCD graph $G$ contains all clades with non-zero weight and all clade splits that can be formed by them. More precisely, if $\cC(G)$ contains clades $C, C_1, C_2$ with $C_1 \cup C_2 = C$ and $C_1 \cap C_2 = \emptyset$, then $\set{C_1, C_2}$ is in $\cS(G)$. The clade weights are converted into clade split probabilities such that the probability of a tree is proportional to the product of its clades' weights \citep{berling25tractable}. For a CCD~$D$, the resulting \emph{clade model probability}~$\prob_D(C)$, the probability that a tree drawn from~$D$ contains clade~$C$, generally differs from the clade weight~$w(C)$.

The parameters of a CCD1 are the non-zero clade split probabilities. A CCD2 extends a CCD1 by distinguishing each clade based on its sibling clade; the probability of a clade split is thus conditional on its parent clade split. CCD2 thus acts on pairs of clade splits, equivalent to the sDAG model \citep{zhang18generalizing, jun23, berling25tractable}.

To populate the parameters of a CCD, we use the frequencies of clades and clade splits in a set of trees $\cT$ coming from an MCMC sample. Let the function~$f$ give the frequency of a clade/clade split appearing in $\cT$. Then for a clade $C$ in a CCD0, the clade weight is set to $w(C) = f(C) / \abs{\cT}$, i.e.\ the Monte Carlo frequency. In a CCD1, the probability of a clade split $\set{C_1, C_2}$ of a clade $C$ is set to $f(\set{C_1, C_2}) / f(C)$, or equivalently, $\prob(\set{C_1, C_2} \mid C)$. In a CCD2, this is further distinguished by the sibling $C'$ of $C$, i.e. $\prob(\set{C_1, C_2} \mid C, C')$. The more complex CCD1 and CCD2 thus require more and more trees to estimate their parameters well. Other methods to populate these parameters are being studied \citep{zhang18variational,jun23,baele24}.

\paragraph{Bias-variance trade-off.}
The sample of trees is another model of the posterior distribution where each tree is a parameter. While this model can perfectly represent the posterior sample, there is no statistical inference of the underlying distribution from which the sample is obtained. In practice, the super-exponential growth of tree-space prevents sampling all trees with non-negligible probability, so, by itself, the posterior sample is generally a poor predictor of individual tree probabilities. The CCDs, on the other hand, employ an independence assumption that clades in separate regions of the tree evolve independently (to some degree). A CCD graph thus contains all combinations of subtrees. Hence, CCDs offer a bias-variance trade-off for estimating the target posterior distribution, namely, the simpler model CCD0 has a higher bias but lower variance, while the more complex model CCD2 has lower bias but higher variance. In our recent study \citep{berling25tractable}, we provided evidence of this trade-off, for both point estimates and estimation of the posterior probability of trees. We found that for non-trivial inference problems, in order to estimate the parameters of a CCD1 (or even a CCD2) to a degree that the CCD1 offers a better point estimate than CCD0, a very large number of sampled trees is required.

\paragraph{MAP Tree.}
For $i \in \set{0, 1, 2}$ and a CCD$[i]$ $D$, the \emph{CCD$[i]$-MAP tree} is the tree~$T$ with maximum probability~$\prob(T)$ in $D$. It can be computed with a dynamic program on the CCD graph with a running time linear in the size of the CCD~graph.

\subsection{Credible Sets} 
\label{sec:credi}
An \emph{$\alpha$ credible set} $C_\alpha$ is a subset of the elements in a discrete distribution~$D$ that together make up at least $\alpha$ of the probability, chosen according to some criterion.
The \emph{credible level}~$\alpha$ lies in $(0, 1]$ and is usually expressed as percentage. For an element $x$ in $C_\alpha$, we say that the \emph{credible level} of $x$ is within $\alpha$; for $x \not\in D$, the credible level of $x$ can be considered infinite.
In general, there are multiple choices for a credible set and a common approach is to find one that contains the fewest elements (the highest posterior density (HPD) credible set). However, given the vast size and complex shape of posterior distributions of phylogenetic trees, this will be computationally intractable in general. The methods described below aim at containing the most probable trees while remaining computationally tractable. 

For any type of credible set of phylogenetic trees, we would like to be able to perform the following tasks efficiently:
\setlist[description]{font=\normalfont\itshape\space}
\begin{description}
	\item[Construction.] Let $D$ be a tree distribution, e.g.\ a sample-based distribution or a CCD.
	For a given credible level $\alpha$ or a range of credible levels, construct the respective credible set(s) of $D$.
	\item[Containment.] Determine whether a tree $T$ is in $C_\alpha$.
	\item[Credible level.] Compute the minimum credible level that contains $T$. 
	\item[Sampling.] Sample a tree from $C_\alpha$ proportional to its probability.
\end{description}
Desirable properties are that the credible sets are \emph{nested}, that is, $C_\alpha \subseteq C_{\alpha'}$ for $\alpha < \alpha'$, and that $C_\alpha$ contains the most probable tree for all $\alpha > 0$. 

We now describe the different approaches to compute credible sets for a range of credible levels $\alpha_1, \ldots, \alpha_\ell$. These $\alpha_i$s represent the granularity of the (pre)computed results, for example, in steps of 1\%. 
For the rest of this section, let $\cT$ be a multi-set of $m$ phylogenetic trees on the same taxa (e.g.\ an MCMC sample) and let $D$ be a CCD based on $\cT$. Recall that $n$ denotes the number of taxa and it takes $\Oh(n)$~time to compute a hash value of a tree. Further let $\abs{D}$ denote the size of $D$ (number of vertices and edges).

\subsubsection{Frequency-based Credible Set.} 
We start with the frequency-based method to compute a credible set from $\cT$ as it is the current default method. As $\alpha_i$ we may take a given range of credible levels (e.g., steps of 1\%) or use credible levels induced by the unique trees in $\cT$; we assume now that we chose the former.

	\noindent\textit{Construction.}
First, sort the different trees in $\cT$ based on their frequency (or equivalently, their Monte Carlo probability) in decreasing order; store the resulting rank of each tree. Then go through the list of sorted trees and add up their Monte Carlo probabilities. Whenever the sum meets or exceeds the next credible level $\alpha_i$ at a tree $T$, we get that $C_{\alpha_i}$ contains all trees up to $T$ and record the rank and probability of $T$ for $C_{\alpha_i}$. These serve as highest rank or lowest probability for a tree to be in $C_{\alpha_i}$. Since there are ties in Monte Carlo probabilities, in particular in the tail, we can alternatively say that $C_{\alpha_i}$ contains all trees with probability at least $\prob(T)$.
Note that the resulting credible sets~$C_{\alpha_i}$, $i \in [\ell]$, are nested.
We can find all unique sampled trees with a hash set in $\Oh(m n)$ time and then sort them in $\Oh(m \log m)$ time;
hence construction takes $\Oh(mn + m \log m)$ time.

	\noindent\textit{Containment.}
We can test containment of a tree $T$ in some $C_{\alpha_i}$ by checking whether the stored probability $\prob(T)$ (or its rank) is at least (resp. at most) the threshold of $C_{\alpha_i}$.
If we first have to find $T$ among the unique sampled trees, say, with a hash set, this adds $\Oh(n)$ time to the otherwise constant running time.

	\noindent\textit{Credible level.}
To determine the credible level of a given tree $T$ (with respect to the $\alpha_i$s), find the smallest $\alpha_i$ such that the probability (rank) stored by $C_{\alpha_i}$ is smaller (resp. greater) or equal to that of $T$. With a binary search, this takes $\Oh(\log \ell)$ time.

	\noindent\textit{Sampling.}
To sample from $C_{\alpha_i}$, we can use rejection sampling: repeatedly sample a tree from $\cT$ (or the unique sampled trees) and accept it if its credible level is at most $\alpha_i$. Since $C_{\alpha_i}$ contains exactly $\alpha_i$ of the probability mass, the rejection rate is $1 - \alpha_i$. 

	\noindent\textit{Remarks.} 
As discussed above, while the frequency-based method works well for trivial datasets with narrow posteriors, the number of ties in Monte Carlo probabilities increases significantly for diffuse posteriors up to the point where all sampled trees are unique. Another major disadvantage is that we cannot determine the credible level of an unsampled tree as its probability is assumed zero by this method.

\subsubsection{Probability-based Credible Set.} 
Our first new method is based on probability thresholds for elements in $C_{\alpha_i}$ obtained by sampling trees from a CCD~$D$. It yields implicit credible sets since we have no explicit representation (data structure) of exactly the trees in a credible set.

	\noindent\textit{Construction.}
Let $k$ be some integer, say, $k = \num{10000}$. (We discuss this choice in the Supplementary Information.) Sample $k$ trees  $T_1, \ldots, T_k$ from the CCD~$D$ and compute the probability $\prob(T_j)$ for each (using~$D$). Then reindex the trees in decreasing order of their probability, so that $\prob(T_1) \geq \prob(T_2) \geq \ldots \geq \prob(T_k)$.
Then, for $j = \ceil{\alpha_i \cdot k}$, let $p_j = \prob(T_j)$ and set $p_j$ as the probability threshold for any tree to be contained in $C_{\alpha_i}$. For example, consider $j = \ceil{k/2}$ and the tree $T_j$ in the middle of the sequence with probability $p_j$. Note that the sample gives evidence that the probability for a random tree in $D$ having a probability higher than $p_j$ is $0.5$. This is not in conflict with there being more trees in the tail with lower probability than $p_j$; we just sampled less of them. Also note that it is possible that $p_j = p_{j + 1}$ as a high probability tree can be sampled multiple times. 
The construction can be done in $\Oh(k \abs{D} + k \log k)$ time since both sampling a tree and computing its probability can be done in $\Oh(\abs{D})$~time. Note that the credible sets are again nested by design.

	\noindent\textit{Containment.} 
By computing $\prob(T)$ of a tree $T$, we can test whether it contained in a specific $C_{\alpha_i}$ in $\Oh(\abs{D})$ time.

	\noindent\textit{Credible level.}
To compute the credible level of a given tree $T$:
\begin{enumerate}
	\item Compute the probability $\prob(T)$ of $T$ in~$D$, in $\Oh(\abs{D})$ time.
	\item Find the smallest credible level $\alpha_i$ whose threshold satisfies $\prob(T) \geq p(\alpha_i)$, via binary search in $\Oh(\log \ell)$ time.
\end{enumerate}
The credible level of $T$ is then within $\alpha_i$, determined in $\Oh(\abs{D} + \log \ell)$ time overall. Note that the probability thresholds are non-increasing in~$\alpha$ (higher credible levels include more trees, so the threshold is lower), which is why a tree with high probability may meet the threshold for multiple credible levels and we take the smallest.

	\noindent\textit{Sampling.}
As with the frequency-based credible set, we can again use rejection sampling for conditional sampling from $C_{\alpha_i}$. Sampling a tree from $D$ can be done in $\Oh(n)$ time and the rejection rate is again $1 - \alpha_i$.

\subsubsection{Clade- and Clade Split-based Credible Sets.} 
Next we describe a method to obtain a credible set represented by a CCD by iteratively removing vertices from the CCD graph of~$D$. Consistent with the CCD model, we either remove clades (CCD0) or clade splits (CCD1, CCD2). In general, for $\alpha \in (0, 1]$, we say a CCD $D_\alpha$ is an \emph{$\alpha$ credible CCD} with respect to $D$ if it contains at least $\alpha$ of the probability mass of $D$. We explain the construction for the CCD0 model, then discuss the changes needed for the CCD1 model.

\noindent\textit{Construction for CCD0.}
Let $D$ be a CCD0. The main idea of the construction is that we repeatedly remove the clade with the lowest clade model probability from $D$ until we are left with only a single tree. We can either explicitly construct and store an $\alpha$ credible CCD $D_\alpha$ or use implicit values (as with the methods above) to later determine credible levels of trees. 
Let $D_0 = D$ and, for $i \geq 0$, let~$D_i$ be the CCD obtained after $i$ iterations. Let $C_i$ be the clade with lowest clade model probability, $\prob_{D_i}(C_i)$, in $D_i$, which can be found in $\Oh(\abs{D})$ time \citep{berling25tractable}. If $\prob_{D_i}(C_i) = 1$, then $D_i$ only contains one tree and we are done. Otherwise, to obtain $D_{i+1}$ from $D_i$, we remove $C_i$ and its neighbourhood as follows:
\begin{enumerate}
  \item Let $C = C_i$.
  \item Remove each clade split $S$ containing $C$; if $S$ was the only clade split of the parent/sibling clade $C'$, then mark $C'$ for deletion.
  \item Remove each clade split $S$ of $C$; if $S$ was the only parent clade split of a child clade $C'$, then mark $C'$ for deletion. 
  \item Remove $C$.
  \item If there is a clade $C'$ marked for deletion, repeat with $C = C'$. 
\end{enumerate}
Note that the continued removal of clades whose only parent or only child gets removed, ensures that the resulting graph is a CCD graph. This process never removes a leaf or the root clade. Note that this would imply that $C$ is contained in every tree and would thus have clade model probability 1; this would contradict the choice of $C$. Also note that all clades in one iteration necessarily have the same clade model probability as $C_i$. Removing the vertices from the CCD graph takes time linear in the number of removed vertices. Therefore, the overall running time for removals is in $\Oh(\abs{D})$ time. Once done with the removal step, we apply the normalization step from the construction of a CCD0 to renormalize all probabilities. In particular, since the trees containing $C_i$ are no longer represented by $D_{i+1}$, the clade model probability in $D_{i+1}$ of any clade (and clade split) that co-occurred with $C_i$ (or another removed clade) decreases, while the model probability of clades that never co-occurred with $C_i$ increases accordingly. Renormalizing takes $\Oh(\abs{D})$ time and, hence, the overall running time is in $\Oh(\abs{D}^2)$.

Let $\prob_\Sigma(D_i)$ represent the probability mass remaining in $D_i$. For $C_i$ and each clade removed alongside, we store that their credible level is $\prob_\Sigma(D_i)$. For $D_{i + 1}$, this value is given by the following formula, since we remove exactly $\prob_{D_i}(C_i)$ probability with respect to $D_i$:
\begin{equation}
	\prob_\Sigma(D_{i + 1}) =  \prob_\Sigma(D_{i}) \cdot \big(1 - \prob_{D_i}(C_i)\big)
\end{equation}
Note that although multiple clades may be removed alongside $C_i$, the total probability mass removed equals exactly $\prob_{D_i}(C_i)$, since every co-removed clade only appears in trees that also contain $C_i$.

\noindent\textit{Construction for CCD1 \& CCD2.}
If $D$ is a CCD1 (or CCD2), then there are three differences in the construction. First, instead of picking a clade $C_i$, we pick the clade split $S_i$ with lowest probability in~$D_i$. Next, when removing $S_i$ to obtain $D_{i+1}$, there is not necessarily any other clade to remove. However, when $S_i$ is the only parent or child clade split of clade~$C$, then as before, we mark $C$ for deletion. If any clade was marked for deletion, we use the algorithm from above to clean up the CCD graph. Finally, and this is the main difference with CCD0, we have to propagate the change in probability upwards, namely, for any clade and clade split on paths from the root clade to $S_i$, we have to account for not choosing a path that leads to $S_i$ anymore. This can be done efficiently in $\Oh(\abs{D})$ time by sorting all clades by size and then, for a clade~$C$, summing up the conditional clade probabilities of all clade splits of $C$. If they add up to 1, there is nothing to do; otherwise, if they add up to $p < 1$, we (i) renormalize the conditional clade probabilities of the clade splits of $C$ and (ii) decrease the conditional clade probabilities of all parent clades splits of $C$ by multiplying them with $p$. We can then recompute the probability of each clade and each clade split in the final $D_{i+1}$.

	\noindent\textit{Containment.}
If we have stored $D_i$, we can test whether a tree $T$ is contained in $D_i$ with existing algorithms, which takes at most $\Oh(\abs{D})$ time. For the implicit method, we simply compute the credible level of $T$ (as follows) and compare it to $\alpha_i$. The explicit method might be of interest if we are interested in a particular credible level, say $\alpha = 0.95$, and also want to sample from $D_i$.

	\noindent\textit{Credible level.}
We can compute the credible level of a tree $T$ by taking the maximum credible level of each clade (clade split for CCD1) contained in $T$. This takes at most $\Oh(n^2)$ time as we have to look up $\Oh(n)$ clades/clade splits, which takes $\Oh(n)$ time each; in practice we may assume though that the look up runs in constant time for the range of $n$ typically encountered.

	\noindent\textit{Sampling.}
It is straightforward to perform restricted sampling directly from $D_i$; alternatively, we could again use rejection sampling from the full CCD without explicitly storing $D_i$.

\noindent\textit{Remarks.} A few notes on the credible CCDs and construction:
\begin{itemize}
  \item For brevity, we refer to this approach as the \emph{clade/split-based credible set method}, and to the credible sets it produces as \emph{credible CCDs}.
  \item The granularity of the credible levels depends on the number of clades or clade splits in the original CCD. Therefore, a CCD1 offers a more fine grained analysis than a CCD0. Furthermore, unlike the frequency- and probability-based methods, credible CCDs naturally assign credible levels to individual clades and clade splits, enabling targeted hypothesis testing at a finer-grained level than whole trees.
  \item A CCD0 $D'$ obtained by continuous removal of clades is equivalent to building a CCD0 $D''$ using only the clades of $D''$ to begin with. However, the equivalent does not hold for CCD1: While a CCD0 already normalizes ``globally" over the whole CCD, a CCD1 only does so locally among all clade splits under each clade~\citep{berling25tractable}.
  \item Here we proposed the minimum probability to pick the next clade or clade split to remove. Alternative strategies are possible 
such as choosing the clade or clade split that is contained in the most trees. Such an alternative strategy can also be used to resolve potential ties.
\end{itemize}

\subsubsection{Computational Complexity Summary.} 
Recall that we let $n$ denote the number of taxa, $m = \abs{\cT}$ the number of sampled trees, $\abs{D}$ the size of the CCD graph (number of vertices and edges), and, if applicable, $\ell$ the number of provided credible levels. Since each of the $m$ trees contributes at most $n - 1$ clades and $n - 1$ clade splits, the number of distinct clades and clade splits is at most $\Oh(mn)$, and the CCD graph has $\abs{D} \in \Oh(mn)$. In practice, $\abs{D}$ is often much smaller due to shared clades across trees.
For example, in our Yule simulations with $n \in \set{10, 20}$ and the real datasets (details below), the CCD graphs contained between $2n$ and $5n$ distinct clades ($3n$ and $13n$ clade splits), far below the $\Oh(mn)$ in the worst case.

We report the asymptotic running times for the four tasks across the different approaches in \cref{tbl:runtimes}. In summary, the dominant cost is the one-time construction of the credible sets, after which queries are fast. All methods scale linearly or quadratically in $\abs{D}$ and are readily applicable to datasets with hundreds of taxa and tens of thousands of sampled trees.
\begin{table}[ht]
    \centering
    \caption{Running times for the different tasks across the credible set methods.}
    \label{tbl:runtimes}
    \begin{tabular}{p{2.5cm}p{3cm}p{3cm}p{3cm}}
        \toprule
        Task 			& Frequency-based & Probability-based & Clade/split-based \\
        \midrule
        Construction    & $\Oh(mn + m\log m)$ & $\Oh(k\abs{D} + k\log k)$  & $\Oh(\abs{D}^2)$ \\
        Containment     & $\Oh(1)$ / $\Oh(n)$ & $\Oh(\abs{D})$             & $\Oh(\abs{D})$ \\
        Credible level  & $\Oh(\log \ell)$    & $\Oh(\abs{D} + \log \ell)$ & $\Oh(n^2)$ \\
        Sampling 		& $\Oh(1)$            & $\Oh(n)$                   & $\Oh(n)$ \\
        \bottomrule
    \end{tabular}
\end{table}

Regarding space requirements, the frequency-based method stores all unique trees explicitly, requiring $\Oh(mn)$ space. The probability-based method requires only $\Oh(\ell + \abs{D})$
space: $\ell$ probability thresholds plus the CCD used to evaluate tree probabilities. The clade/split-based CCD method similarly requires $\Oh(\ell \cdot \abs{D})$ for the explicit version (storing a pruned CCD per credible level) or $\Oh(\abs{D})$ for the implicit version (the CCD and one float per clade/split). In practice, since $\abs{D}| \ll mn$, both CCD-based methods are considerably more compact than the frequency-based approach for diffuse posteriors with many unique trees.

\subsection{Model Evaluation \& Datasets}
\label{sec:evaluation}
We demonstrate two applications of credible sets: evaluating the performance of well-calibrated simulation studies (WCSS) \citep{mendes2024validate} and assessing the goodness of fit for different CCD models. Additionally, we compare the effectiveness of the different credible set methods. We use WCSS datasets derived from simple models extensively tested in BEAST2 and generated with LinguaPhylo \citep{drummond23linguaphylo}, which are expected to pass a coverage analysis. Therefore, they provide a reliable basis for evaluating CCD models. In particular, we use the following WCSS datasets from \citet{berling25tractable}, shortly summarized as follows:
\begin{itemize}[leftmargin=*,label=XXXXX]
  \item[\texttt{Yule[$n$]}] For number of taxa $n \in \set{50, 100, 200}$, we used 250 Yule tree \citep{yule1925ii} simulations, each with birth-rate prior having a log-normal distribution $(\ln 25, 0.3)$, HKY+G substitution model \citep{hasegawa1985dating}, shape parameter with gamma distribution of site rates using log-normal distributions (${-}1.0$, $0.5$), transition/transversion rate ratio $\kappa$ with log-normal distribution ($1.0$, $1.25$), nucleotide base frequencies independently sampled for each simulation from a Dirichlet distribution with a concentration parameter array of [5.0, 5.0, 5.0, 5.0], sequence alignments with 300 sites, and a fixed mutation rate at $1.0$, leading to branches in units of expected substitutions per site.
  \item[\texttt{Coal[$n$]}] For each number of taxa $n \in \set{40, 80, 160}$, we used 250 time-stamped coalescent \citep{rodrigo1999coalescent} simulations, each with population size ($\theta$) with log-normal distribution (${-}$2.4276, 0.5), sequence alignments with 250 sites, with youngest leaf at age 0 and the remaining leaf ages uniformly at random between 0 and 0.2, and with all other parameters as in the Yule simulations. 
\end{itemize}
The differing numbers of taxa in the Yule and coalescent models were intentionally selected to yield more comparable entropy profiles across the datasets.

In addition, the following datasets have phylogenetic entropies of around 5 and a small number of taxa. For these we obtained \emph{golden runs} of \num{100000} trees that represent the posterior distributions sufficiently well enough (except in the tails) such that the frequency-based credible sets can be used as a ground truth.
\begin{itemize}[leftmargin=*,label=XXXXX]
  \item[\texttt{Yule[$n$]}] We use the same parameters as for the Yule simulations above with $n \in \set{10, 20}$, changing only to fixed birth rates (25.0 for $n = 10$ and 12.5 for $n = 20$) and increasing the number of sites to 600 for $n = 20$.
  \item[\texttt{DS1-4}] These are the four smallest datasets from a suite of popular benchmarking datasets in phylogenetics \citep{lakner2008efficiency} 
  with the number of taxa being 27, 29, 36, and 41. We use BEAST2 replicates from \citet{berling24automated}.
  The dataset DS1 is believed to have a bimodal posterior tree distribution \citep{whidden2015quantifying}.
\end{itemize}
For the cross-model comparison (\cref{sec:results:cephalopod}), we used three empirical datasets from \citet{douglas2025evolution}, each analysed under both a gradual and a gradual+abrupt clock model:
\begin{itemize}[leftmargin=*,label=Indo-Europeanx]
  \item[\texttt{aaRS}] A curated alignment of 494 amino acid sites from 142 Class~I aminoacyl-tRNA synthetase (aaRS) catalytic domains, sourced from bacteria, archaea, eukaryotes, and viruses.
  \item[\texttt{Cephalopods}] A morphological dataset of 153 discrete traits (shell shapes, tentacle structures, fins, and other properties) for 27 living and 52 fossil cephalopod taxa, with the oldest fossil (\emph{Nautilus pompilius}) dated at 469\,Ma.
  \item[\texttt{Indo-European}] A dataset of \num{4990} cognates across 109 present-day and 52 ancient Indo-European languages, with the oldest languages (Hittite, Luvian, Mycenaean Greek) dating to before 1000\,BCE.
\end{itemize}
For each simulation and each empirical dataset, we used an MCMC sample of \num{10000} trees for evaluation (besides the golden runs). Note that the \texttt{Yule[$n$]} datasets have contemporaneous taxa while the \texttt{Coal[$n$]} datasets have serially-sampled~taxa.

\subsubsection{Sensitivity \& Specificity.}
To evaluate the different credible set methods, we use the trivial datasets \texttt{Yule10}, \texttt{Yule20}, and \texttt{DS1-4}. Using the golden runs as a baseline, we compare the credible level of each sampled tree within each CCD[$i$] model, for $i \in \set{0,1,2}$, and a sample distribution on \num{10000} trees. In particular, we compute the sensitivity and specificity (explained below) of each method. Corresponding plots depict sensitivity and specificity against credible level thresholds to illustrate how well each model captures the posterior distribution. These plots enable visual comparisons of model performance and help identify systematic biases or limitations in specific CCD models or the posterior sample model.

Recall that the sensitivity is defined as the ratio of true positives over the number of elements within the credible set and the specificity is defined as the ratio of the true negatives over the number of elements outside the credible set. A high sensitivity implies that most of the elements included in the credible set are true positives, meaning the method captures the relevant trees well. However, this does not ensure that the right trees are excluded, as false positives may still be present unless specificity is also high. Conversely, a high specificity means that most of the elements excluded from the credible set are true negatives, indicating that the method effectively filters out the right trees. High specificity but not so great sensitivity means that the method tends to include only trees it is very confident about, at the cost of failing to capture many true positives. This results in a credible set that is precise but incomplete.

\subsubsection{True Tree Coverage Analysis.}
Suppose that we have decided which CCD model and which credible set method to use, e.g.\ CCD1 and the probability-based method. Now given a WCSS dataset with $k$ simulations, we test the coverage of the true tree as follows. For each simulation, with true tree $T_0$ and MCMC sample~$\cT$, construct the CCD1 $D$. Then test whether $T_0$ lies in the 95\% credible set of $D$. Assuming that this is true for $k'$ of the simulations, report the coverage ratio $\frac{k'}{k}$. Furthermore, we say that the WCSS passes the coverage test for the tree topology if $k'$ lies in the 95\%-central interquantile interval of a binomial distribution; for $\alpha = 0.95$ and $k = 100$, this is between 90 and 99 (see \citealt{mendes2024validate} for details).
Note that unlike for HPD intervals of continuous parameters, we cannot readily visualize the credible set and the placement of $T_0$ within a credible set. 

A more extensive evaluation, also allowing for diagrams, follows the idea of rank-uniformity validation. For each of the $k$ simulations, compute the credible level of $T_0$ within $D$. Then group the credible levels into buckets of size 1\% (or your preferred choice), i.e. buckets $[0\%, 1\%)$, $[1\%, 2\%)$, $\ldots$, $[99\%, 100\%]$. As a result we get a frequency distribution, which we can visualize using a histogram.

For a more global assessment of goodness of fit, \citet{sailynoja22} suggest to use the \emph{Empirical Cumulative Distribution Function (ECDF)}. The ECDF represents cumulative counts over the buckets, normalised by $k$, and can be plotted alongside 95\% simultaneous confidence intervals based on the binomial distribution. These intervals provide a statistical benchmark to evaluate whether the observed cumulative counts align with the expected uniform distribution under a well-calibrated model.

We provide examples of both the histogram and the ECDF diagrams in the CCD Model Evaluation section, where we use the WCSS datasets to compare both the CCD models and the credible set methods.

\section{Results} 
\label{sec:results}
We first evaluate the different credible set methods on trivial datasets. Then we demonstrate how the credible set methods work for WCSS and evaluate the different CCD models.

\subsection{Credible Set Method Evaluation} \label{sec:results:eval}
\begin{figure}[ht]
  \centering
  \includegraphics[width=\linewidth]{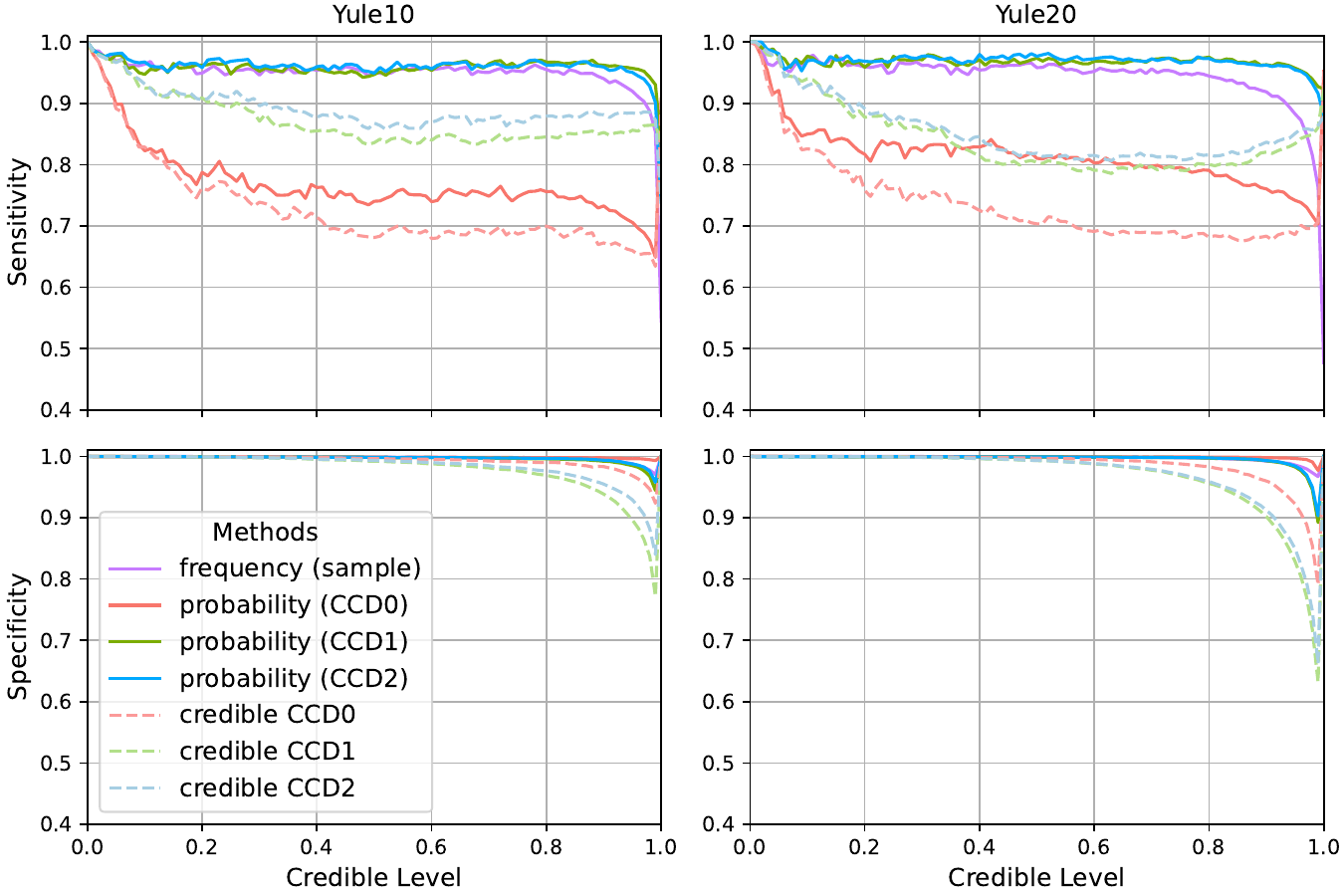}
  \caption{Mean sensitivity and specificity of the different credible set methods over 250 replicates of \texttt{Yule10} and \texttt{Yule20}. (Note that the y-axes start at $0.4$.)}
  \label{fig:sensi:yule:tenk}
\end{figure}

\begin{figure}[ht]
  \centering
  \includegraphics[width=\linewidth]{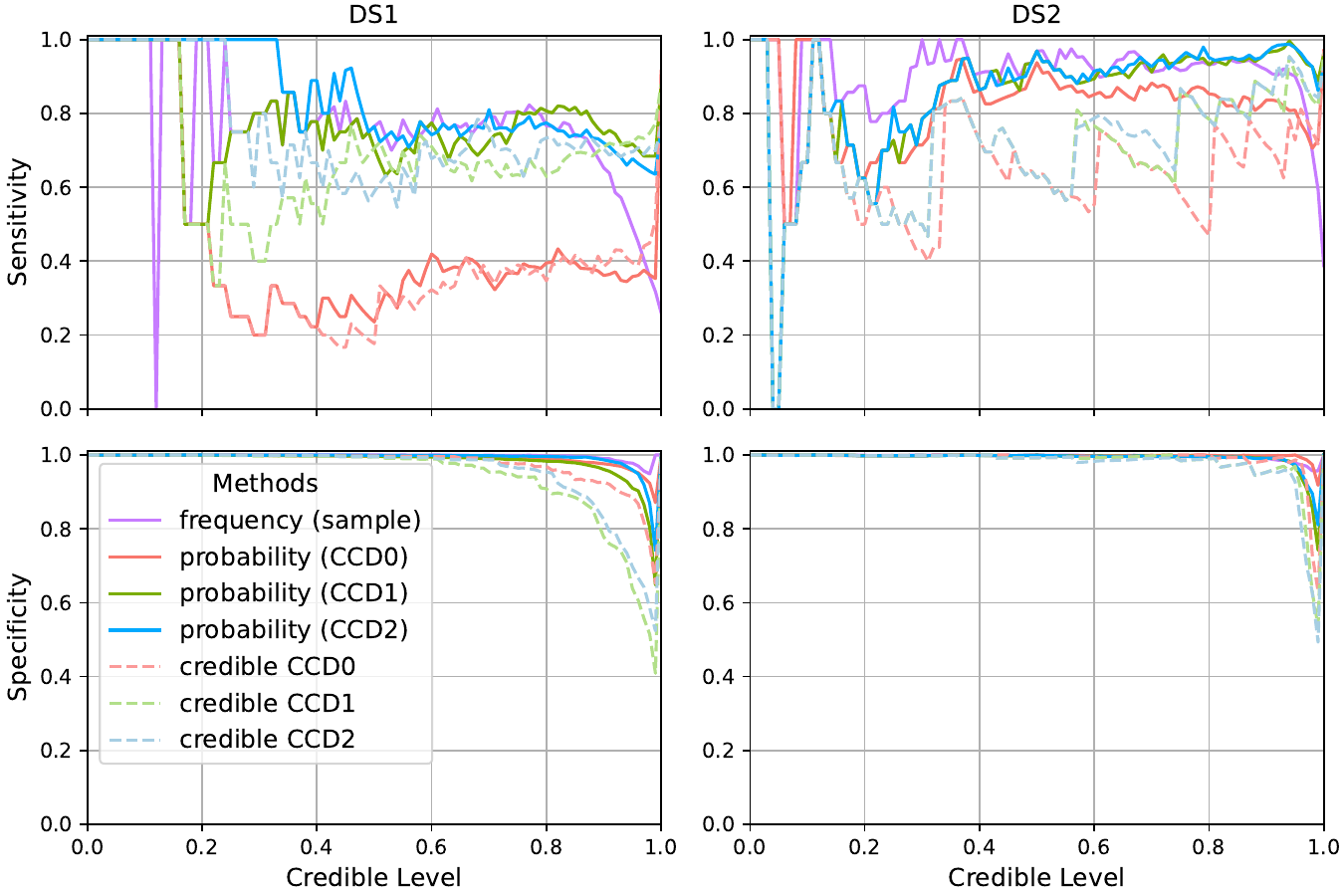}
  \caption{Sensitivity and specificity of the different credible set methods on \texttt{DS1-2}.}
  \label{fig:sensi:ds:tenk}
\end{figure}

We use the golden runs consisting of \num{100000} trees as a baseline. Then we used \num{10000} trees to represent as sample distribution and to construct a CCD with each model (CCD0, CCD1, and CCD2). We evaluated the frequency-based method on the sample distribution, the probability-based method in conjunction with each CCD model (using \num{10000} samples and \num{1000} steps), and each credible CCD model. For each unique tree present in the golden run, we calculated their credible level both in the golden run and using each of the credible set methods. Using credible levels incremented in steps of $0.1\%$ (resulting in \num{1000} steps), we then compute the sensitivity and specificity of each method. The results for \texttt{Yule10} and \texttt{Yule20}, in the form of the mean over the 250 replicates, and for the empirical datasets \texttt{DS1} and \texttt{DS2} are shown in \cref{fig:sensi:yule:tenk,fig:sensi:ds:tenk}. (Additional results for \texttt{Yule10} and \texttt{Yule20} using \num{3000} trees (Fig.~S1) and those for datasets \texttt{DS3} and \texttt{DS4} using \num{3000} and \num{10000} trees (Figs.~S2--S3) and for datasets \texttt{DS1} and \texttt{DS2} using \num{10000} trees (Fig.~S4) are presented in the Supplementary Information.)


Regarding  specificity, all methods performed exceptionally well up to credible levels of 80\% or higher. Beyond this point, the methods begin to include trees into the credible sets that are excluded by the frequency-based credible sets of the golden run. This could be due to tail effects: Trees beyond the 95\% credible level are typically observed only 3--4 times in the golden run, making their frequency-based probability estimates unreliable and the golden run itself an imperfect ground truth in this regime.

Turning to sensitivity, we observe that, particularly for the real datasets, the curves exhibit a jagged pattern in the first half. This is likely due to minor variations in credible levels among the trees with the highest probabilities. We observe that for each type of CCD, the probability-based credible sets perform better than the clade/split-based one. However, methods based on CCD1 and CCD2 demonstrate superior performance compared to those based on CCD0. This indicates that CCD1 and CCD2 capture the shape of the posterior distribution better than CCD0. Furthermore, comparing the \texttt{Yule10} and \texttt{Yule20} results reveals that the probability-based credible sets derived from CCD1 and CCD2 outperform the frequency-based method as the dataset complexity increases. Even on \texttt{DS1}, where the CCDs might have difficulty capturing the bimodal nature of the posterior distribution, their performance is comparable to or better than that of the frequency-based method.

In summary, the high specificity and sensitivity of the probability-based CCD1 and CCD2 credible sets indicate that they are slightly conservative, tending to exclude some trees that are included in the frequency-based credible set of the golden run, but rarely including those that should not. The credible sets using CCDs may result in better credible levels of low probability trees than with the frequency-based method.

\subsection{CCD Model Evaluation} \label{sec:results:model}
\begin{figure}[tbh]
  \centering
  \includegraphics[width=\linewidth]{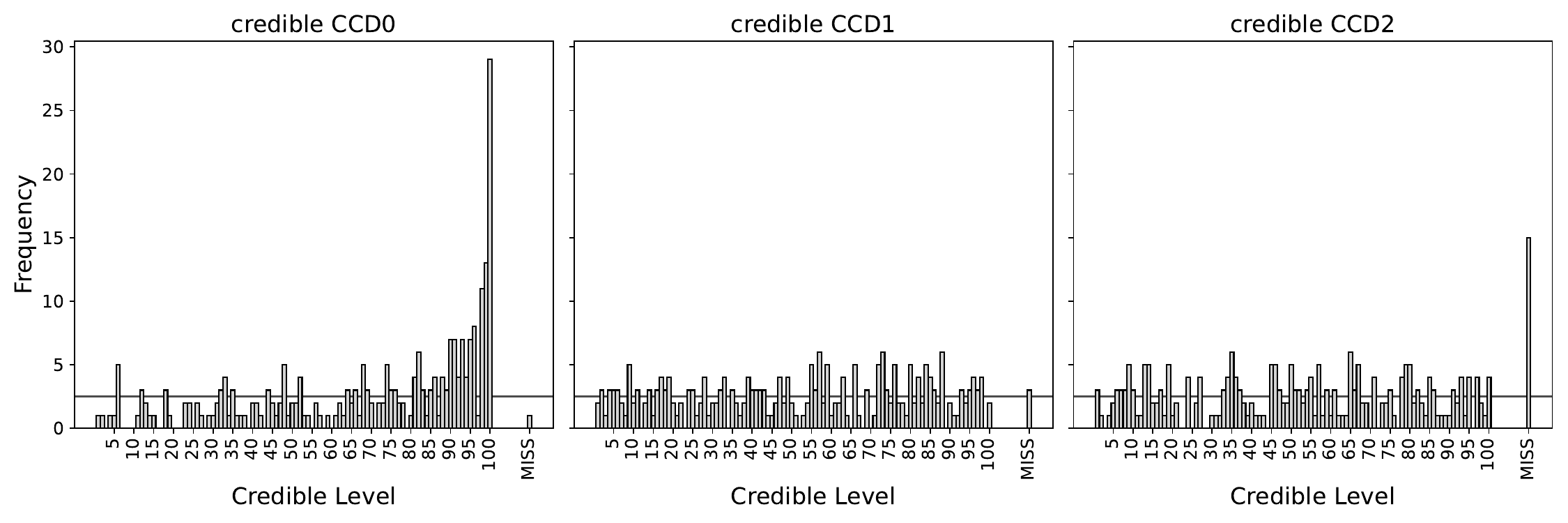}
  \caption{Histograms of the credible levels of the true tree topology with 1\% buckets for the different CCD models and the \texttt{Yule50} dataset.}
  \label{fig:model:histo}
\end{figure}

\begin{figure}[tbh]
  \centering
  \includegraphics[width=\linewidth]{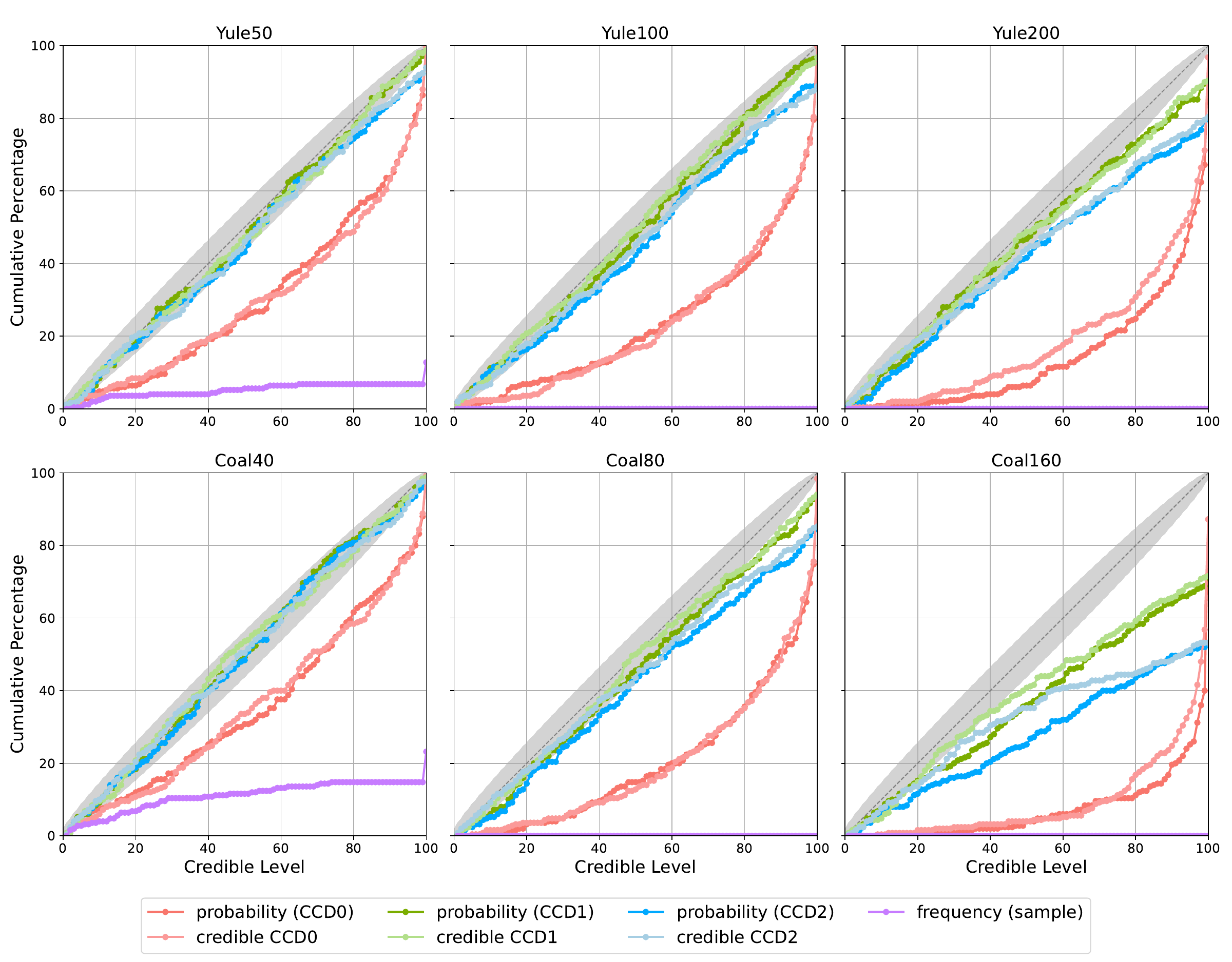}
  \caption{ECDF diagrams for the true tree topology with 1\% buckets for the different CCD models based on an MCMC sample of 10,000 trees as well as the 95\% central interquantile intervals for the different $\alpha$ intervals covering the true parameters (based on binomial distribution with 250 trials and probability of success $\alpha$).}
  \label{fig:model:ecdf}
\end{figure}

To evaluate CCD models with WCSSs, we applied all credible set methods on the datasets \texttt{Yule}[$n$], $n \in \set{50, 100, 200}$, and \texttt{Coal}[$n$], $n \in \set{40, 80, 160}$. Each dataset comprised 250 replicates, and \num{10000} trees were used for the sample distribution and to construct the CCDs. The credible level of the true tree for each replicate was then computed with each method. Histograms depicting the distribution of the credible levels in the clade/split-based credible sets for the \texttt{Yule50} dataset, using a bin width of 1\%, are presented in~\cref{fig:model:histo}. Furthermore, \cref{fig:model:ecdf} presents the Empirical Cumulative Distribution Function (ECDF) plots for all six datasets, showing the proportion of true trees with a credible level less than or equal to a given value (in 1\% increments) for each method, along with the 95\% central interquantile intervals.

As expected, the frequency-based credible set method performed poorly, given that sample distributions rarely contain the true tree (and almost never for larger datasets). This method could assign a credible level to the true tree in only approximately 10\% of the replicates for \texttt{Yule50} and \texttt{Coal40}, and failed entirely for the larger datasets.

Let us consider first the relative performance of the remaining methods. For each CCD type, the probability-based and the clade/split-based credible sets exhibited comparable results, with neither method demonstrating a clear advantage. However, comparing the CCD types shows that CCD1 and CCD2 consistently outperformed CCD0. While CCD1 and CCD2 showed similar performance for the two smaller datasets, CCD2’s performance declined relative to CCD1 as the dataset complexity increased. This could be due to CCD2 requiring more data than CCD1 for good parameter estimation. The steep upward trend of CCD0 for high credible levels, even overtaking CCD1 and CCD2 at 100\%, can be explained by CCD0 containing more trees than the other models and thus also more often the true tree.

To assess the absolute performance, we examine whether the ECDFs fall within the 95\% central interquantile intervals. We observe a clear failure of the CCD0-based methods across all six datasets, generally underestimating the credible level of the true tree. This observation implies that CCD0 may excessively flatten the underlying probability landscape. The CCD1-based methods demonstrated good performance for \texttt{Yule50}, \texttt{Yule100}, and \texttt{Coal40}, with their ECDF curves generally falling within the 95\% confidence interval. However, for the larger datasets, a slight underestimation of the true tree’s credible level was observed. The CCD2-based methods exhibited slightly more underestimation, extending to the \texttt{Yule100} dataset as well. Thus, CCD1 and CCD2 may not fully capture the shape of the larger datasets when constructed with only \num{10000} samples. (Considering the effective sample size (ESS) of the sample may also show that the MCMC sample showed higher levels of auto-correlation for the larger datasets.) Furthermore, they seem to struggle more with the time-stamped data in the Coalescent datasets.


\subsection{Application: Cross-model Comparison} \label{sec:results:cephalopod}

We used credible sets to compare posterior tree topology distributions inferred under two competing evolutionary clock models, a gradual model and a gradual+abrupt model \citep{douglas2025evolution}, on three empirical datasets: a cephalopod morphological dataset (80 taxa), an aminoacyl-tRNA synthetase (aaRS) protein structural dataset (142 taxa), and an Indo-European language dataset (161 taxa). Following \citet{berling25tractable}, we used CCD0 for point estimation (MAP trees) and CCD1 for split-based credible sets. The CCD0 MAP trees differ in topology under all three datasets (rooted Robinson--Foulds distances of 18/156, 22/280, and 12/318, respectively). However, in all cases the MAP trees lie well within the CCD1 credible set of the opposing model (credible levels ranging from 0.00 to 0.30). Recall that the credible level of a tree within a distribution is the probability mass of the smallest credible set containing it, so lower values indicate that the tree is among the most probable trees in the opposing model's posterior. This demonstrates that the topological differences are not statistically significant. The RF distance alone cannot tell us whether such differences are significant given the posterior uncertainty; credible sets can. See Supplementary Information for full details, including the effect of the number of sampled trees on credible sets (Fig.~S5) and cross-model comparison results for all three empirical datasets (Table~S1).

\section{Discussion} 
\label{sec:discussion}
We introduced novel methods to compute credible sets on phylogenetic tree topologies for Bayesian analyses using MCMC. The probability-based credible sets are straightforward to implement and can be easily adapted to other tree distributions. The credible CCDs go beyond representing just a credible set by offering a full tractable tree distribution, from which can be sampled and to which other algorithms can be applied. We further showed that credible CCDs can be constructed efficiently. The performance of all methods was evaluated on trivial datasets against a frequency-based credible set on golden runs. Furthermore, we demonstrated the utility of these methods in assessing the fit of CCD models to posterior distributions of tree topologies.

Traditional frequency-based credible sets, used since the earliest Bayesian phylogenetic analyses \citep{mau1997phylogenetic,mau1999bayesian,yang1997bayesian}, are limited to trees present in the MCMC sample and cannot assign credible levels to unsampled trees. For diffuse posteriors where most or all sampled trees are unique, this approach assigns equal probability to every sampled tree and the resulting credible set becomes essentially arbitrary. Our CCD-based methods overcome both limitations: by leveraging the independence structure of CCDs, they assign probabilities to all trees representable by the observed clades and clade splits, including the vast majority that were never sampled. Similarly, while frequentist confidence sets based on likelihood ratio tests \citep{goldman2000likelihood,shimodaira2002approximately} provide an alternative form of uncertainty quantification, they require per-topology likelihood evaluation and are typically applied to small candidate sets of competing hypotheses, making them impractical for characterising the full posterior landscape. In contrast, our methods scale to the full support of a CCD, which may contain orders of magnitude more topologies than the MCMC sample. Furthermore, the clade/split-based approach uniquely provides credible levels for individual clades and clade splits, enabling finer-grained hypothesis testing than any whole-tree method.

Our experiments comparing credible set methods revealed a slight advantage in favour of the probability-based credible set over the credible CCDs, regardless of the underlying CCD model. This could be due to the probability-based method conceptually aligning more closely with the frequency-based method. Notably, we did not observe a worse performance of the clade/split-based credible sets in the CCD model evaluation with ECDF plots. This suggests that clade/split-based credible sets remain a viable alternative. The strong similarity observed between these methods, particularly on larger datasets, lends further confidence to their reliability. Credible CCDs produced by the clade/split-based approach further offer the distinct advantage of assigning credible levels to individual clades and clade splits, enabling targeted hypothesis testing at a finer-grained level than for whole trees. However, for analyses where such detailed clade-specific information is not required, the probability-based method may be preferred due to its inherent simplicity.

Our evaluation of CCD models yielded several key insights. CCD0 consistently failed to accurately capture the shape of the posterior distributions, while CCD1 demonstrated good performance on the smaller datasets. We also observed reduced performance on the time-stamped Coalescent datasets compared to the contemporaneous Yule datasets. It is important to note that these credible set results contrast with our earlier findings \citep{berling25tractable}, where CCD0 generally provided the superior point estimate of the tree topology based on the Robinson-Foulds distance.

These results show that the quality of CCD-based credible sets is bounded by the quality of the underlying CCD as a model of the posterior distribution. CCDs rely on an independence assumption, that clade splits in different parts of the tree are conditionally independent, which may be violated when there are strong correlations between distant clades. When this assumption is inadequate, the CCD will mis-estimate tree probabilities, and the resulting credible sets may have poor coverage. Furthermore, current CCD models focus solely on topological structure without incorporating temporal information, which may partially explain the reduced performance on time-stamped data. Future development of CCD models should prioritize the integration of such temporal data. The choice of CCD model also involves a bias-variance trade-off that depends on the sample size: CCD1 and CCD2 require large numbers of effectively independent samples to estimate their parameters well, and there are currently no automatic guidelines for model selection, though our credible set methods can themselves be used to evaluate model fit. Additionally, the greedy removal strategy used to construct credible CCDs is a heuristic that does not guarantee the smallest possible credible set (i.e.\ the HPD set), and the probability-based method depends on a stochastic sampling step whose variability decreases with the number of samples $k$ (cf.\ Supplementary Information). We also observed tail effects in the posterior samples, even in our golden runs on small datasets; these are expected to set in once differences in clade split probability estimates fall within the Monte Carlo sampling error, at which point the CCD can no longer reliably rank the trees it assigns low probability. To mitigate these effects, future work could explore pre-filtering tree samples by removing clades and clade splits observed only once, or applying regularization techniques to CCD models. The WCSS datasets used in our experiments represent simplified scenarios that may not fully encompass the complexities inherent in real-world phylogenetic posterior distributions, though the failure of CCD0 even in these experiments strongly suggests an underlying lack of modelling capacity.

Despite these limitations, the CCD1- and CCD2-based credible sets demonstrated remarkably good calibration on the datasets studied, with ECDF curves largely falling within the 95\% confidence intervals for the smaller datasets.

Lastly, while our methods and experiments here focused on rooted phylogenetic trees, there is no inherent obstacle to adopt them to tractable distributions of unrooted phylogenetic trees.

\section{Conclusion} 
\label{sec:conclusion}
Our novel credible set methods for phylogenetic tree topologies provide new capabilities for phylogenetic analyses, particularly in cases with diffuse posteriors where the frequency-based method fail. These methods allow targeted hypothesis testing for phylogenetic trees or parts thereof, as well as rank uniformity tests and ECDF plots in WCSS. We hope that by enabling more accurate and reliable quantification of uncertainty in tree topology estimates, our methods can lead to more robust inferences in downstream analyses.

Making use of these new capabilities, we evaluated the performance of the CCDs models on simulated datasets. We found that CCD1 and CCD2 remaining promising models but CCD0 showed a clear lack of modelling capacity. We hope that our approach enables further development and evaluation of tractable tree distributions. 
In particular, we would like to see CCDs that incorporate temporal information. The probability-based credible set method could be readily adapted to such a distribution.

\section*{Data Availability} 
An open source implementation is freely available in the CCD package for BEAST2 hosted on GitHub: \href{https://github.com/CompEvol/CCD}{\texttt{github.com/CompEvol/CCD}}.
The empirical datasets used in the cross-model comparison are from \citet{douglas2025evolution} and are available on Dryad: \href{https://doi.org/10.5061/dryad.sxksn03dj}{\texttt{doi.org/10.5061/dryad.sxksn03dj}}.

\section*{Acknowledgements} 
JK and AJD were partially supported by the Beyond Prediction Data Science Research Programme (MBIE grant UOAX1932).

\renewcommand{\bibname}{References}
\pdfbookmark[1]{References}{References}
\bibliographystyle{mbe-natbib}
\bibliography{sources}

\clearpage
\appendix
\pdfbookmark[0]{Supporting Information}{Appendix}
\section*{Supporting Information}
\section{Additional Sensitivity and Specificity Analyses}
\begin{figure}[h]
  \centering
  \includegraphics[width=\linewidth]{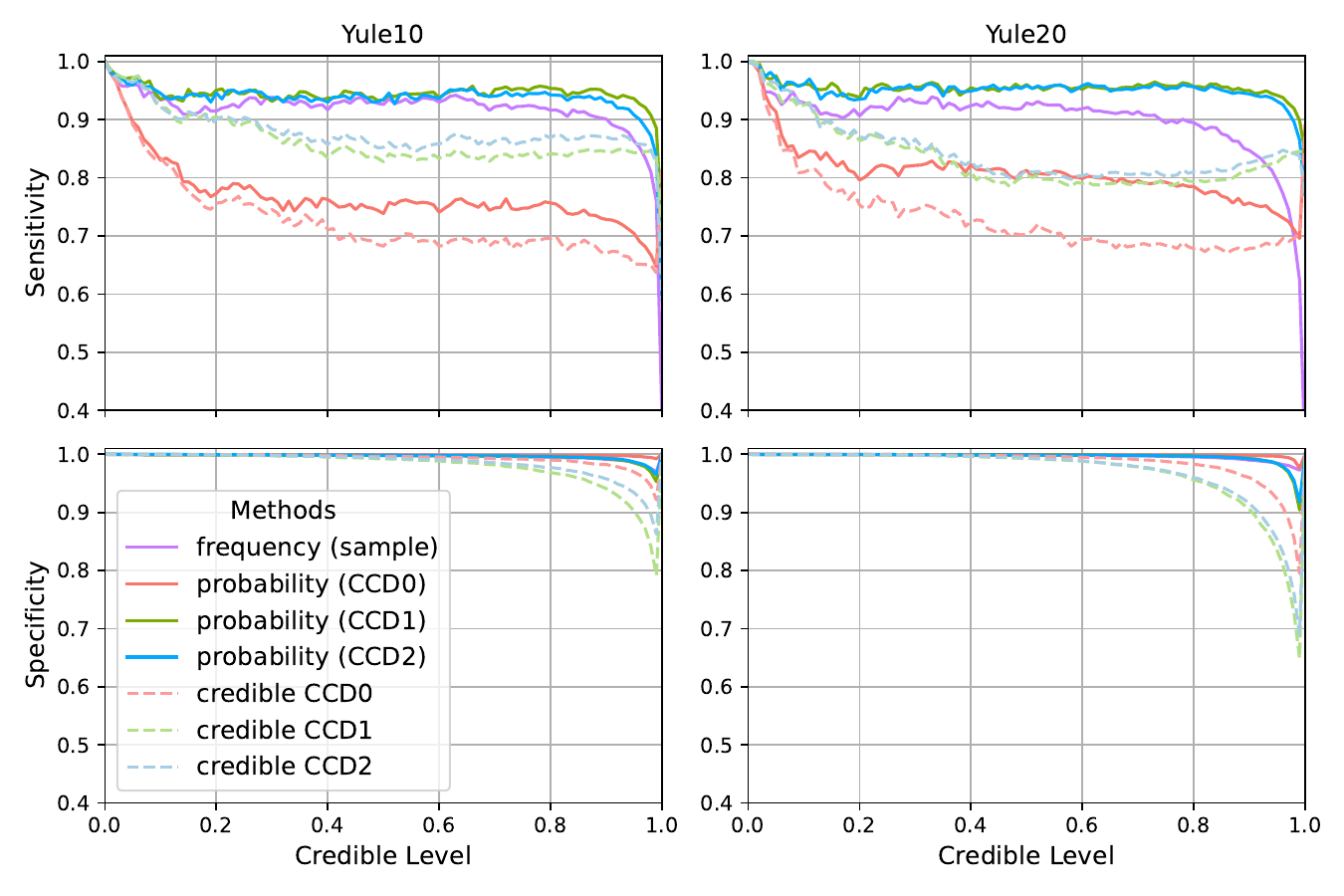}
  \caption{Mean sensitivity and specificity of the different credible set methods over 250 replicates of \texttt{Yule10} and \texttt{Yule20} with \num{3000} trees used in the sample distribution and to build the CCDs. (Note that the y-axes start at $0.4$.)}
  \label{fig:sensi:yule:threek}
\end{figure}

\begin{figure}[ht]
  \centering
  \includegraphics[width=\linewidth]{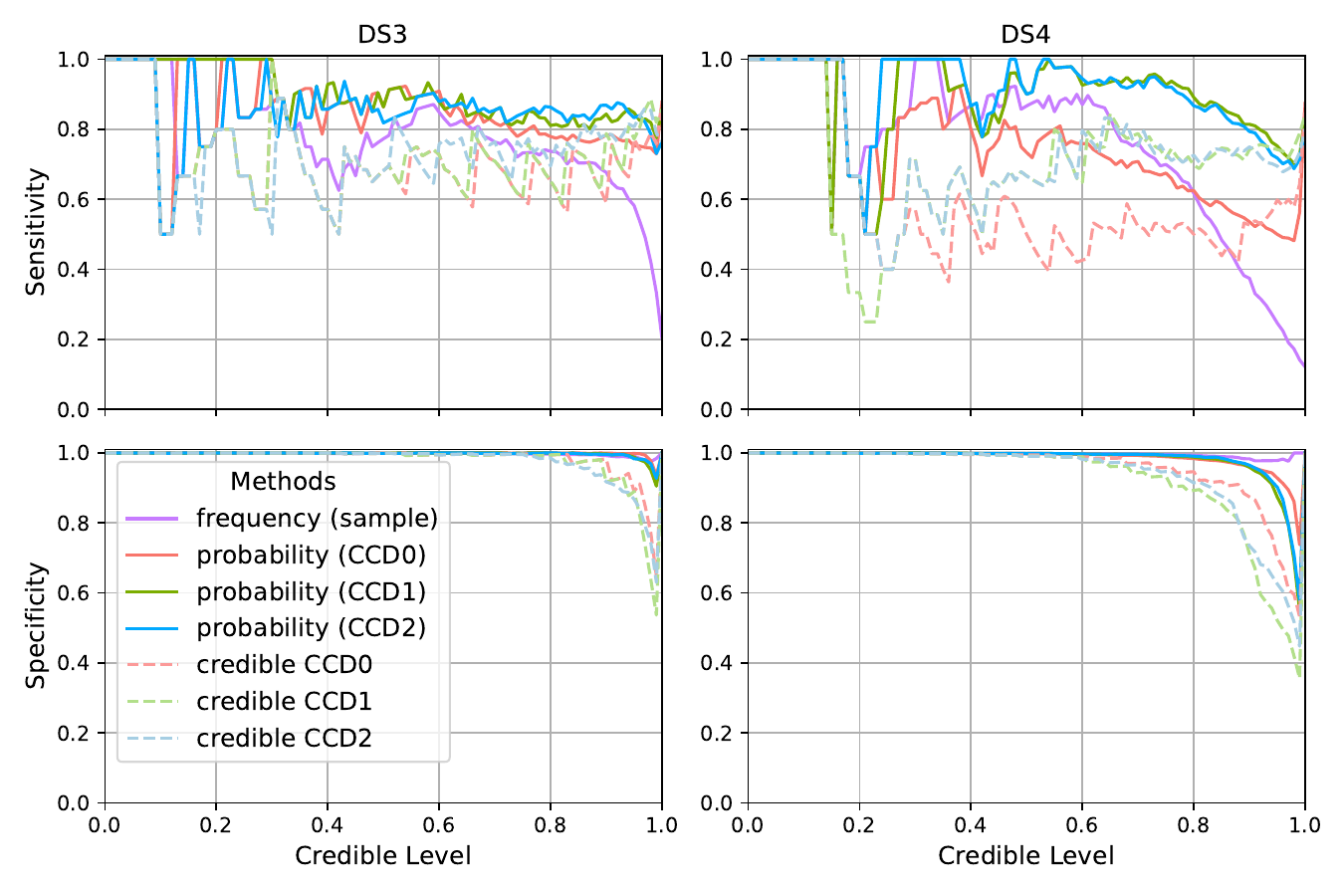}
  \caption{Sensitivity and specificity of the different credible set methods on \texttt{DS3-4} with \num{3000} trees used in the sample distribution and to build the CCDs.}
  \label{fig:sensi:ds:threek}
\end{figure}

\begin{figure}[ht]
  \centering
  \includegraphics[width=\linewidth]{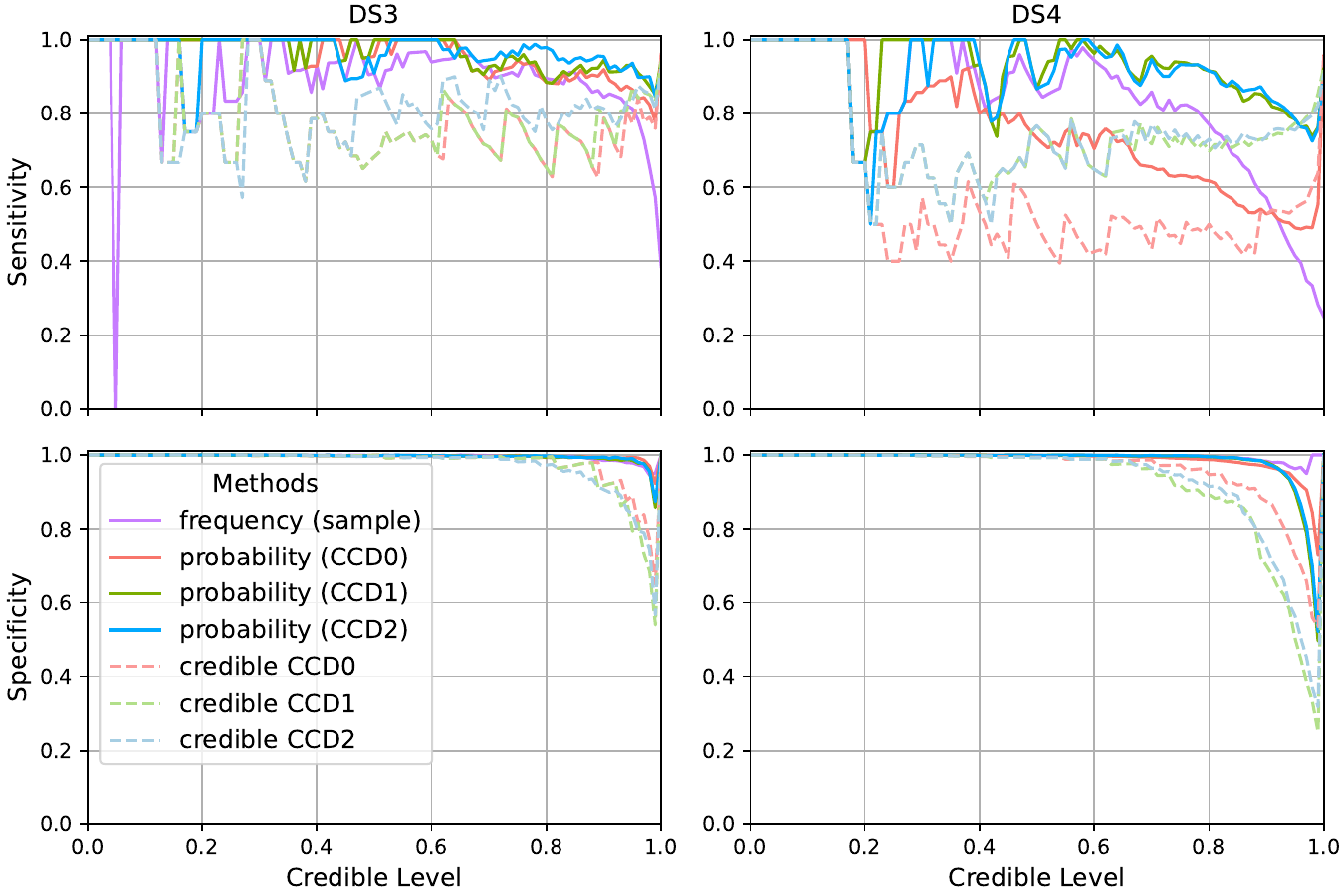}
  \caption{Sensitivity and specificity of the different credible set methods on \texttt{DS3-4} with \num{10000} trees used in the sample distribution and to build the CCDs.}
  \label{fig:sensi:ds:other:tenk}
\end{figure}

\begin{figure}[ht]
  \centering
  \includegraphics[width=\linewidth]{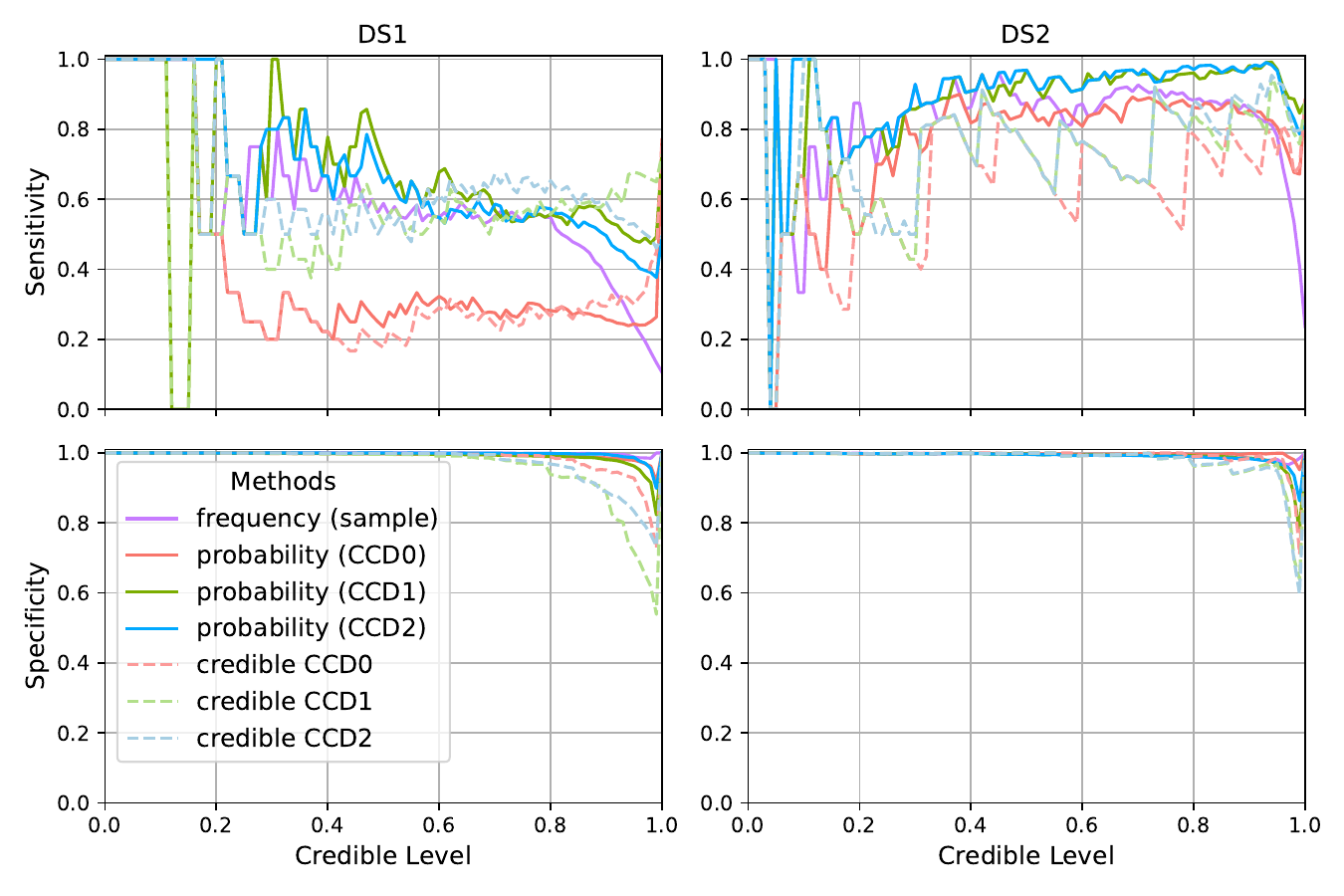}
  \caption{Sensitivity and specificity of the different credible set methods on \texttt{DS1-2} with \num{10000} trees used in the sample distribution and to build the CCDs.}
  \label{fig:sensi:ds:another:tenk}
\end{figure}

We provide additional sensitivity and specificity results in \cref{fig:sensi:yule:threek,fig:sensi:ds:threek,fig:sensi:ds:other:tenk,fig:sensi:ds:another:tenk}.

\FloatBarrier

\section{Probability-based Method Parameter Choice} \label{app:parachoice}
\begin{figure}[tbh]
  \centering
  \includegraphics[width=\linewidth]{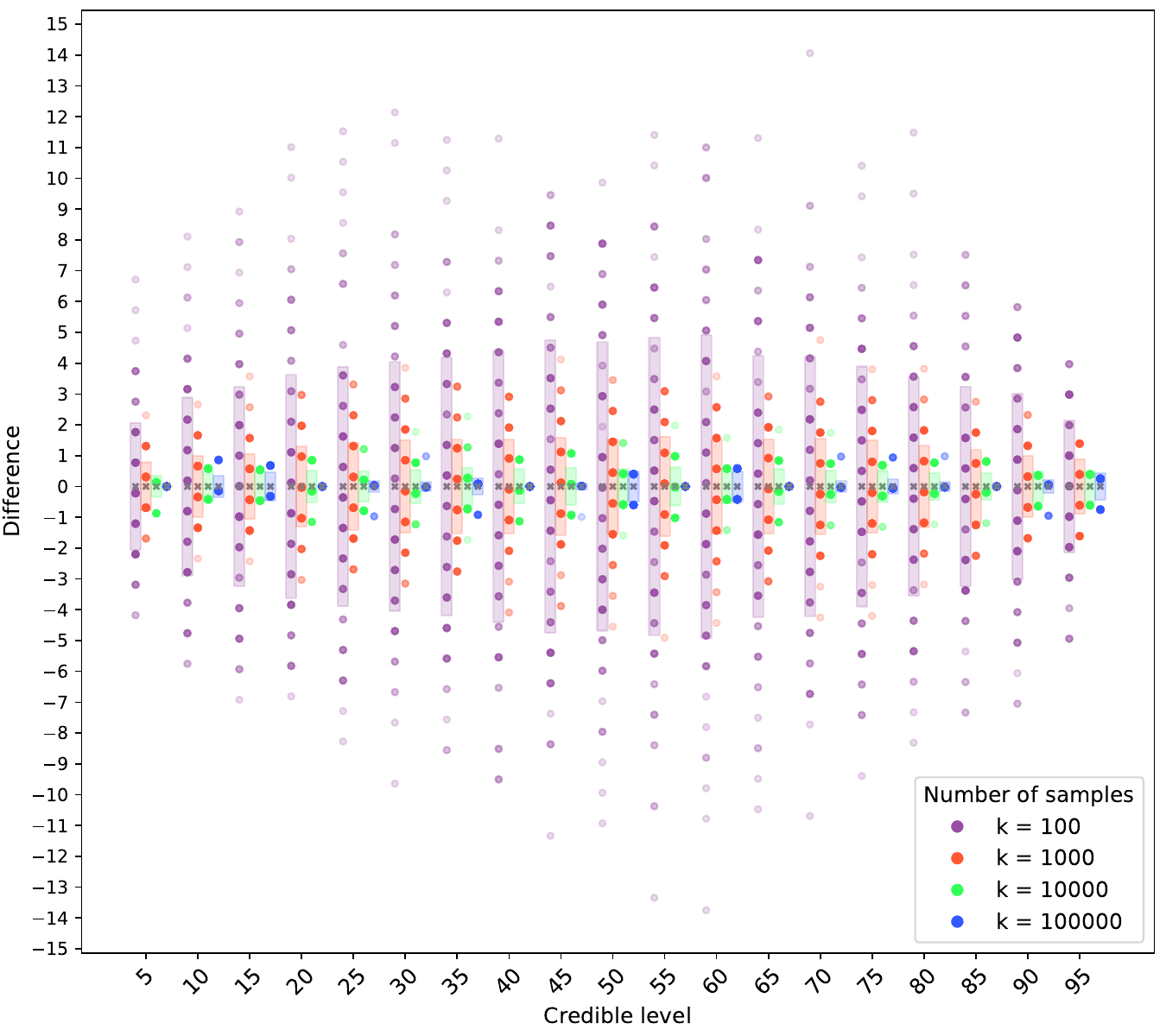}
  \caption{Difference in the computed credible level of random trees (one per multiple of 5\%)
  by the probability-based credible set method on a CCD1 when using different numbers of samples.}
  \label{fig:parachoice}
\end{figure}

We want to demonstrate the effect of the sample size used for probability-based credible sets with a step size of 1\%, here with CCD1 and using \num{100}, \num{1000}, \num{10000}, and \num{10000} samples on a \texttt{Yule50} simulation. To this end, we sampled random trees with a credible level that is a multiple of 5 from a probability-based CCD1 credible set on \num{100000} samples. Then computed credible sets (the probability thresholds) for the different number of samples one hundred times. The difference to mean for each sample size is illustrated in \cref{fig:parachoice}. We conclude that, at least for a \texttt{Yule50} dataset, \num{10000} samples yield an acceptably small variance. 

\FloatBarrier

\section{Cross-Model Comparison on Empirical Data} \label{app:cephalopod}

To illustrate a practical application of credible sets, we compared posterior tree topology distributions inferred under two competing evolutionary clock models \citep{douglas2025evolution}: a gradual clock model and a gradual+abrupt model that additionally allows for instantaneous bursts of evolution at speciation events.
We analysed three datasets from that study: a cephalopod morphological dataset (80 taxa), an aminoacyl-tRNA synthetase (aaRS) protein structural dataset (142 taxa), and an Indo-European language dataset (161 taxa).
Following \citet{berling25tractable}, we used CCD0 to obtain point estimates (MAP trees) under each model, and CCD1 to construct split-based credible sets.

In all three datasets, the CCD0 MAP trees of the two clock models differ in topology, with rooted Robinson--Foulds distances of 18/156 (cephalopod), 22/280 (aaRS), and 12/318 (Indo-European).
A natural question is whether these topological differences are statistically significant given the posterior uncertainty around each tree.
The RF distance alone cannot answer this question, but credible sets can.
As shown in \cref{tbl:crossmodel}, all MAP trees lie well within the CCD1 credible set of the opposing model: the highest credible level observed is 0.30 (aaRS gradual MAP tree in the gradual+abrupt credible set), meaning that tree is still more probable than 70\% of all trees in that posterior.
In no case does a MAP tree approach the 95\% credible set boundary.
The topological differences between the two clock models are therefore not statistically significant in any of the three datasets.

\citet{douglas2025evolution} noted that the cephalopod clade support differences between the two clock models were ``not as extreme'' as those observed for the aaRS dataset, and that the aaRS models ``produced substantially different clade posterior supports''.
Our credible set analysis quantifies these qualitative observations at the whole-tree level: while all three datasets show non-significant topological differences, the aaRS MAP trees sit further from the core of the opposing credible set (credible levels up to 0.30) compared to the cephalopod trees (up to 0.08) and the Indo-European trees (up to 0.002).
Importantly, even for the aaRS dataset where individual clade support differences were statistically significant, the overall tree topologies are not significantly different.

\begin{table}[h]
  \centering
  \caption{Cross-model credible set analysis. CCD0 MAP trees are used as point estimates; CCD1 split-based credible levels quantify how deep each MAP tree sits in the opposing model's posterior. Lower values indicate the tree is closer to the core of the credible set; a value approaching 1.0 would indicate the tree lies near the boundary. Trees used after 10\% burn-in: cephalopod 8551/4009, aaRS 1622/1907, Indo-European 1440/4074.}
  \label{tbl:crossmodel}
  \begin{tabular}{llcc}
    \toprule
     & & \multicolumn{2}{c}{CCD1 credible set of} \\
    \cmidrule(lr){3-4}
    Dataset & CCD0 MAP tree from & gradual & gradual+abrupt \\
    \midrule
    \multirow{2}{*}{Cephalopod (RF = 18/156)}
      & gradual         & 0.00 & 0.08 \\
      & gradual+abrupt  & 0.00 & 0.00 \\
    \midrule
    \multirow{2}{*}{aaRS (RF = 22/280)}
      & gradual         & 0.00 & 0.30 \\
      & gradual+abrupt  & 0.01 & 0.00 \\
    \midrule
    \multirow{2}{*}{Indo-European (RF = 12/318)}
      & gradual         & 0.00 & 0.00 \\
      & gradual+abrupt  & 0.00 & 0.00 \\
    \bottomrule
  \end{tabular}
\end{table}

\end{document}